\documentclass{ws-ijmpa}
\usepackage{amsmath,amssymb,graphicx}

\begin{document}

\title{Flavoured CP asymmetries for type II seesaw leptogenesis}

\author{R. Gonz\'{a}lez Felipe}
\address{Instituto Superior de Engenharia de Lisboa - ISEL,
Rua Conselheiro Em\'{\i}dio Navarro,\\ 1959-007 Lisboa, Portugal}
\address{Departamento de F\'{\i}sica and Centro de F\'{\i}sica Te\'{o}rica de Part\'{\i}culas (CFTP),
Instituto Superior T\'{e}cnico, Universidade de Lisboa, Avenida Rovisco Pais,
1049-001 Lisboa, Portugal\\ricardo.felipe@ist.utl.pt}

\author{F.~R.~Joaquim}
\address{Departamento de F\'{\i}sica and Centro de F\'{\i}sica Te\'{o}rica de Part\'{\i}culas (CFTP),
Instituto Superior T\'{e}cnico, Universidade de Lisboa, Avenida Rovisco Pais,
1049-001 Lisboa, Portugal\\filipe.joaquim@ist.utl.pt}

\author{H.~Ser\^{o}dio}
\address{Departament de F\'{\i}sica Te\`{o}rica and IFIC, Universitat de Val\`{e}ncia - CSIC,\\
E-46100, Burjassot, Spain\\hugo.serodio@ific.uv.es}

\maketitle

\begin{abstract}
A novel contribution to the leptonic CP asymmetries in type II seesaw
leptogenesis scenarios is obtained for the cases in which flavour effects
are relevant for the dynamics of leptogenesis. In the so-called flavoured
leptogenesis regime, the interference between the tree-level amplitude of
the scalar triplet decaying into two leptons and the one-loop wave-function
correction with leptons in the loop, leads to a new nonvanishing CP
asymmetry contribution. The latter conserves total lepton number but
violates lepton flavour. Cases in which this novel contribution may be
dominant in the generation of the baryon asymmetry are briefly discussed.

\keywords{Leptogenesis, neutrino physics, seesaw mechanism}
\end{abstract}

\section{Introduction}

Leptogenesis\cite{Fukugita:1986hr} is perhaps the most appealing mechanism to
explain the matter-antimatter asymmetry observed in the Universe. One of its
remarkable features is the possibility of establishing a bridge between
neutrino physics at high and low energies, through the well-known seesaw
mechanism for neutrino mass generation. Several scenarios are conceivable in
this context. Namely, the canonical ones are the type
I\cite{Minkowski:1977sc,Yanagida:1979as,GellMann:1980vs,Glashow:1980,Mohapatra:1979ia},
type
II\cite{Konetschny:1977bn,Mohapatra:1980yp,Cheng:1980qt,Lazarides:1980nt,Schechter:1980gr}
and type III\cite{Foot:1988aq} seesaws, in which neutrino masses are mediated
by the three-level exchange of  heavy singlet fermions, $SU(2)$-triplet
scalars and $SU(2)$-triplet fermions, respectively. Particularly economical
is the type II seesaw scenario with one triplet, where the flavour pattern of
the Yukawa couplings between the scalar triplets and the Standard Model (SM)
doublets uniquely determines the flavour structure of the low-energy
effective neutrino mass matrix. This feature is particularly interesting when
the type II seesaw is embedded in a beyond-the-SM framework, where those
Yukawa couplings trigger new sources of lepton flavour violation that may be
relevant for processes like radiative charged-lepton
decays\cite{Branco:2011zb}. This is indeed what happens in the supersymmetric
type II seesaw where model-independent predictions can be made for the rates
of lepton flavour violating decays in terms of the low-energy neutrino
parameters\cite{Rossi:2002zb,Joaquim:2006uz,Joaquim:2006mn,Esteves:2009vg,Joaquim:2009vp,Brignole:2010nh}.

To successfully implement leptogenesis in a minimal type II seesaw framework
(without introducing heavy singlet fermions), at least two scalar triplets
are needed\cite{Ma:1998dx}.\footnote{In the presence of only one scalar
triplet, the CP asymmetry induced by the triplet decays is generated beyond
the one-loop level and is therefore highly suppressed\cite{Hambye:2003rt}.}
The complex Yukawa couplings of the Higgs triplets to leptons, as well as
their complex couplings to the standard model Higgs doublet, provide the
necessary sources of CP violation for leptogenesis.  In particular, the CP
asymmetry in the decay of the scalar triplet into two leptons arises from the
interference of the corresponding tree-level and one-loop amplitudes. A
nonvanishing lepton asymmetry is then generated via the out-of-equilibrium
decays of the triplet scalars in the early Universe, which is afterwards
partially converted into a baryon asymmetry by nonperturbative sphaleron
processes\cite{Klinkhamer:1984di}.

Departure from thermal equilibrium crucially depends on the expansion rate of
the Universe. Since at very high temperatures ($T \gtrsim 10^{12}$~GeV) all
charged lepton flavours are out of thermal equilibrium, their states are
indistinguishable. Interactions involving the $\tau$ and $\mu$ Yukawa
couplings enter in equilibrium at $T \lesssim 10^{12}$~GeV and $T \lesssim
10^9$~GeV, respectively. The corresponding lepton doublets are
distinguishable mass eigenstates below these temperature scales and,
therefore, their flavour effects should be properly taken into account in the
leptogenesis dynamics. Such effects turn out to be relevant in type I seesaw
leptogenesis
scenarios\cite{Barbieri:1999ma,Abada:2006fw,Nardi:2006fx,Abada:2006ea}. In
particular, in the flavoured regime the washout processes can be less
significant than in the unflavoured one, and the low-energy leptonic phases
affect directly the final asymmetry so that it is possible to have successful
leptogenesis just from low-energy leptonic CP
violation\cite{Branco:2006ce,Pascoli:2006ie}. Also, the upper bound on each
individual flavoured asymmetry is not suppressed when the absolute neutrino
mass scale increases.

So far, flavour effects in type II seesaw leptogenesis have been only
partially addressed\cite{Blanchet:2008zg,Branco:2011zb,Branco:2012vs}. The
purpose of this work is to study the importance of these effects on the
leptonic CP asymmetries generated in minimal type II seesaw scenarios where
leptogenesis is implemented through the out-of-equilibrium decays of scalar
triplets into two leptons. The required CP asymmetries in those decays are
guaranteed by the interference of the tree-level and one-loop decay
amplitudes, in the presence of complex couplings of the triplets with the SM
Higgs and the leptons. It turns out that there is a novel contribution to the
flavoured leptogenesis asymmetries coming from the wave-function
renormalization correction. We briefly discuss some cases in which this
contribution may be dominant.

\section{Type II seesaw leptogenesis}

In the type II seesaw mechanism, neutrino masses are generated through the
tree-level exchange of hypercharge $Y=1$ scalar SU(2)$_L$ triplets. As
already pointed out, a single triplet is enough to explain the low-energy
neutrino spectrum. However, in this case it is not possible to generate a
leptonic CP asymmetry for leptogenesis, since all the interference terms
vanish at one loop. Thus, we need to add at least another triplet to allow
for the generation of a non-zero CP asymmetry\cite{Ma:1998dx}. Here, we will
consider an extension of the SM in which $n_L$ scalar  triplets $T_a$ are
added. Following the usual SU(2) representation, each $T_a$ can be written in
terms of the corresponding charge eigenstates $T_a^0,T_a^{+}$ and $T_a^{++}$
as
\begin{figure}[t]
\begin{center}
\includegraphics[width=12cm]{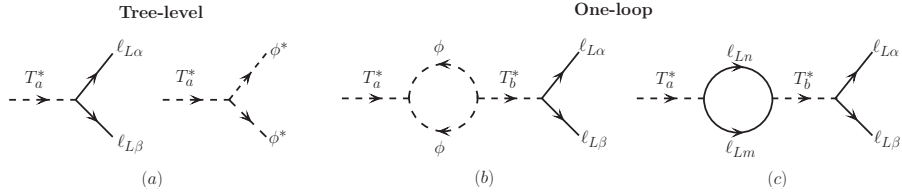}
\end{center}
\caption{\label{fig1} Tree-level and one-loop Feynman diagrams contributing
to the CP asymmetry in a type II seesaw framework.}
\end{figure}

\begin{equation}\label{scalarT}
T_a=
\begin{pmatrix}
T_a^0&-\dfrac{T_a^{+}}{\sqrt{2}}\\
-\dfrac{T_a^{+}}{\sqrt{2}}&T_a^{++}
\end{pmatrix}\,.
\end{equation}

The relevant Lagrangian is given by

\begin{align}\label{lagraII}
\mathcal{L}_{II}=\mathcal{L}_{SM}+
\text{Tr}\left[\left(D_\mu T_a\right)^\dagger
\left(D^\mu T_a\right)\right]-
\left(\mathbf{Y}^{T_a}_{\alpha\beta}\overline{\ell_{L\alpha}}T^\dagger_a
\ell^{c}_{L\beta}+\text{h.c.}\right)-V_{T}\,,
\end{align}
where $\mathcal{L}_{SM}$ contains the SM terms and $V_{T}$ accounts for
the scalar potential terms involving the triplets,
\begin{align}
\begin{split}
V_{T}=&M_{a}^2\,
\text{Tr}\left(T_a^\dagger T_a\right)+
\mu_a\tilde{\phi}^TT_a\tilde{\phi}
+g_{ab}\phi^\dagger T_a^\dagger T_b\phi
+h_{ab}\phi^\dagger\phi\text{Tr}\left(T_a^\dagger T_b\right)\\
&+\lambda^\prime_{abcd}\text{Tr}\left(T^\dagger_aT_b
T^\dagger_cT_d\right)+\lambda_{abcd}
\text{Tr}\left(T_a^\dagger T_b\right)\text{Tr}\left(T_c^\dagger T_d\right)
+\text{h.c.}\,.
\end{split}
\end{align}
In the above equations, $D_\mu$ stands for the usual covariant derivative,
$\phi=(\phi^+,\phi^0)^T$ is the SM Higgs doublet ($\tilde{\phi} = i \sigma_2
\phi^\ast$), and $\ell=(\nu_L,l_L)^T$ is a SM lepton doublet.

Each $T_a$ has two decay modes: $T_a^\dagger\rightarrow
\ell_\alpha\ell_\beta$ and $T_a^\dagger\rightarrow \phi^\ast\phi^\ast$
(see Fig.~\ref{fig1}a). For the first channel, neglecting the masses of the
final states, the tree-level decay rates are
\begin{align}\label{calphabeta}
\begin{split}
\Gamma\left(T^\dagger_a\rightarrow \ell_{\alpha} \ell_{\beta}\right)&
=\frac{M_a}{8\pi}\left|\mathbf{Y}^{T_a}_{\alpha\beta}\right|^2c_{\alpha\beta}\,,
\quad \text{with}\quad c_{\alpha\beta}=\left\{
\begin{array}{ll}
2-\delta_{\alpha\beta}&\text{for }T^{0}_a\,,\, T^{++}_a\\
1&\text{for }T^{+}_a
\end{array}\right..
\end{split}
\end{align}
Summing over the final flavour states we get
\begin{align}
\begin{split}
\Gamma \left(T_a^{--}\rightarrow l^-l^-\right)=
\Gamma \left(T_a^{0\ast}\rightarrow\nu\nu\right)=
&\frac{M_a}{8\pi}\sum_{\alpha,\beta\ge \alpha}
\left|\mathbf{Y}^{T_a}_{\alpha\beta}\right|^2c_{\alpha\beta}\,,\\
\Gamma \left(T_a^{-}\rightarrow l^-\nu\right)=
&\frac{M_a}{8\pi}\sum_{\alpha,\beta}\left|\mathbf{Y}^{T_a}_{\alpha\beta}\right|^2.
\end{split}
\end{align}
Note that in the case of $T_a^{--}$ and $T_a^{0}$ the sum is
ordered, while for $T_a^-$ is not. Therefore, at tree level $\Gamma
\left(T_a^{--}\rightarrow l^-l^-\right)=\Gamma
\left(T_a^{0\ast}\rightarrow\nu\nu\right)=\Gamma
\left(T_a^{-}\rightarrow l^-\nu\right)$. This is the result one would
expect since $T_a^{0\ast}$, $T_a^{-}$ and $T_a^{--}$ belong to
the same $SU(2)_L$ multiplet, which is not yet broken at the leptogenesis
scale. For the decay channel into the Higgs scalars, we get
\begin{align}
\Gamma \left(T_a^{--}\rightarrow \phi^-\phi^-\right)=\Gamma
\left(T_a^{0\ast}\rightarrow \phi^{0\ast}\phi^{0\ast}\right)=\Gamma
\left(T_a^{-}\rightarrow \phi^-
\phi^{0\ast}\right)=\frac{\left|\mu_a\right|^2}{8\pi M_a}\,.
\end{align}
The total decay rate is then given by
\begin{align}
\Gamma_{T_a}\equiv\Gamma\left(T^\dagger_a\rightarrow \ell \ell,\phi^\ast
\phi^\ast\right)=\frac{M_a}{8\pi}\left[\text{Tr}\left(\mathbf{Y}^{T_a\dagger}
\mathbf{Y}^{T_a}\right)+\left|\lambda_a\right|^2\right]\,,
\end{align}
with $\lambda_a=\mu_a/M_a$.

The leptonic CP asymmetries relevant for leptogenesis stem from the
interference between the tree-level and one-loop $T_a$ decay amplitudes. In
our framework, there is no vertex correction contributing to the leptogenesis
CP asymmetry at the one-loop level. The only diagrams contributing to the CP
asymmetries are those coming from wave function renormalization, shown in
Figs.~\ref{fig1}b and \ref{fig1}c. Notice that the former is both total
lepton number and lepton flavour violating, while the second one is only
lepton flavour violating. The interference between the one-loop diagram (1b)
and the corresponding  tree-level one leads to the CP
asymmetry\cite{Branco:2011zb}
\begin{align}\label{CPasymmIIflavour1}
\begin{split}
\epsilon_a^{\alpha\beta}(\text{wave
1})\simeq&\frac{c_{\alpha\beta}}{2\pi}\frac{\sum_{b\neq
a}g(z_b)\text{Im}\left(\lambda_a^\ast\lambda_b\mathbf{Y}^{T_b}_{\alpha\beta}
\mathbf{Y}^{T_a\ast}_{\alpha\beta}\right)}{\text{Tr}
\left(\mathbf{Y}^{T_a\dagger}\mathbf{Y}^{T_a}\right)
+|\lambda_a|^2}\,,
\end{split}
\end{align}
where $z_b\equiv M_b^2/M_a^2$ and
\begin{equation}\label{gz}
g(z_b)=\frac{\sqrt{z_b}\,(1-z_b)}{(z_b-1)^2 +(\Gamma_{T_b}/M_a)^2}\,.
\end{equation}

We recall that in the type II seesaw the Yukawa couplings are directly
related to the effective light neutrino mass matrix by
\begin{equation}
\label{mnudef}
\mathbf{m}_{\nu}=\sum_{a}\mathbf{m}_{\nu}^a\;,\quad
\mathbf{m}_{\nu}^a=2\langle
T_a^{0}\rangle^\ast\mathbf{Y}^{T_a}=2\frac{\lambda_a v^2}{M_a}\mathbf{Y}^{T_a}\,,
\end{equation}
with $v=\langle \phi^{0}\rangle=174$~GeV. One can then rewrite the leptonic
CP asymmetries in terms of these quantities. Indeed, using the branching
ratio relations
\begin{equation}
\mathcal{B}_a^\ell\, \Gamma_{T_a}\equiv
\frac{M_a}{8\pi}\text{Tr}\left(\mathbf{Y}^{T_a\dagger}\mathbf{Y}^{T_a}\right)\,,\quad
\mathcal{B}_a^\phi\, \Gamma_{T_a}\equiv \frac{M_a}{8\pi}|\lambda_a|^2\,,
\end{equation}
and
\begin{align}
\begin{split}
\sqrt{\mathcal{B}_a^\ell\mathcal{B}_a^\phi}\,\Gamma_{T_a}=&\frac{M_a^2}{16\pi
v^2}\left[\text{Tr}(\mathbf{m}_{\nu}^{a\dagger}
\mathbf{m}_{\nu}^a)\right]^{1/2},\\
\text{Im}\left(\lambda_a^\ast\lambda_b\mathbf{Y}_{\alpha\beta}^{T_b}
\mathbf{Y}_{\alpha\beta}^{T_a\ast}\right)=&
\frac{M_aM_b}{4v^4}\text{Im}\left[\left(\mathbf{m}_{\nu}^b\right)_{\alpha\beta}
\left(\mathbf{m}_{\nu}^{a\ast}
\right)_{\alpha\beta}\right]\,,
\end{split}
\end{align}
Eq.~(\ref{CPasymmIIflavour1}) can be expressed in the form
\begin{equation}
\epsilon_a^{\alpha\beta}(\text{wave
1})\simeq\frac{c_{\alpha\beta}}{4\pi}\frac{M_a\sqrt{\mathcal{B}_a^\ell\mathcal{B}_a^\phi}}{v^2}
\frac{\sum_{b\neq
a}g(z_b)\text{Im}\left[\left(\mathbf{m}_{\nu}^b\right)_{\alpha\beta}
\left(\mathbf{m}_{\nu}^{a\ast}\right)_{\alpha\beta}\right]}{\left[\text{Tr}(\mathbf{m}_{\nu}^{a\dagger}
\mathbf{m}_{\nu}^a)\right]^{1/2}}\,.
\end{equation}
For the second contribution to the wave function renormalization, coming from
the interference between the one-loop diagram (1c) and the corresponding
tree-level one, we obtain
\begin{align}\label{CPasymmIIflavour2}
\epsilon_a^{\alpha\beta}(\text{wave
2})&\simeq\frac{c_{\alpha\beta}}{2\pi}\frac{\sum_{b\neq a} z_b^{-1/2}
g(z_b)\, \text{Im}\left[\text{Tr}\left(\mathbf{Y}^{T_b\dagger}
\mathbf{Y}^{T_a}\right)\mathbf{Y}^{T_b}_{\alpha\beta}
\mathbf{Y}^{T_a\ast}_{\alpha\beta}
\right]}{\text{Tr}\left(\mathbf{Y}^{T_a\dagger}\mathbf{Y}^{T_a}\right)
+\left|\lambda_a\right|^2}\,.
\end{align}
Rewriting the imaginary part as
\begin{equation}
\text{Im}\left[\text{Tr}\left(\mathbf{Y}^{T_b\dagger} \mathbf{Y}^{T_a}\right)
\mathbf{Y}^{T_b}_{\alpha\beta}\mathbf{Y}^{T_a\ast}_{\alpha\beta}\right]
=\frac{M_a^2M_b^2}{16v^8|\lambda_a|^2|\lambda_b|^2}\text{Im}\left[\text{Tr}
\left(\mathbf{m}_{\nu}^{b\dagger} \mathbf{m}_{\nu}^{a}\right)(\mathbf{m}_{\nu}^{b})_{\alpha\beta}
(\mathbf{m}_{\nu}^{a\ast})_{\alpha\beta}\right]\,,
\end{equation}
we get for the CP asymmetry
\begin{align}
\begin{split}
\epsilon_a^{\alpha\beta}(\text{wave 2})\simeq & \frac{c_{\alpha\beta}}{16\pi}
\frac{M_a\sqrt{\mathcal{B}_a^\ell\mathcal{B}_a^\phi}}{v^6|\lambda_a|^2
\left[\text{Tr}\,( \mathbf{m}_{\nu}^{a\dagger}\mathbf{m}_{\nu}^a)\right]^{1/2}}\times\\
&\sum_{b\neq a} \frac{M_b^2\, g(z_b)\,\text{Im}\left[\text{Tr}
\left(\mathbf{m}_{\nu}^{b\dagger} \mathbf{m}_{\nu}^{a}\right)(\mathbf{m}_{\nu}^{b})_{\alpha\beta}
(\mathbf{m}_{\nu}^{a\ast})_{\alpha\beta}\right]}{z_b^{1/2}|\lambda_b|^2}\,,
\end{split}
\end{align}
in terms of the various contributions to the neutrino mass matrix.

The above novel contribution to the CP asymmetries is only relevant within
the flavoured leptogenesis regime. Indeed, summing over the final flavours we
get
\begin{align}
\epsilon_a(\text{wave }2)\propto \text{Im}\left[\text{Tr}\left(\mathbf{Y}^{T_b\dagger}
\mathbf{Y}^{T_a}\right)\text{Tr} \left(\mathbf{Y}^{T_a\dagger}
\mathbf{Y}^{T_b}\right)\right]=0\,.
\end{align}
Therefore, the only contribution surviving in the unflavoured regime comes
from the Higgs loop. The final unflavoured asymmetry is given by
\begin{align}\label{CPasymmIIa}
\begin{split}
\epsilon_a &\simeq\frac{1}{2\pi}\frac{\sum_{b\neq a}g(z_b)
\text{Im}\left[\lambda_a^\ast\lambda_b \text{Tr}\left(\mathbf{Y}^{T_a\dagger}
\mathbf{Y}^{T_b}\right)\right]}{\text{Tr}\left[\mathbf{Y}^{T_a\dagger}
\mathbf{Y}^{T_a}\right]
+|\lambda_a|^2}\\
 &=\frac{1}{4\pi}\frac{M_a\sqrt{\mathcal{B}_a^\ell\mathcal{B}_a^\phi}}{v^2}
 \frac{\sum_{b\neq a}g(z_b)\text{Im}\left[\text{Tr}\left(\mathbf{m}_{\nu}^{a\dagger}
 \mathbf{m}_{\nu}^b\right)\right]}{\left[\text{Tr}\left(\mathbf{m}_{\nu}^{a\dagger}
 \mathbf{m}^a_{\nu}\right)\right]^{1/2}}\,.
\end{split}
\end{align}

We stress that in the regime where the final-state flavour discrimination is
important, the new contribution $\epsilon_a^{\alpha\beta}({\rm wave \,2})$
given in Eq.~(\ref{CPasymmIIflavour2}) can, in principle, dominate over
$\epsilon_a^{\alpha\beta}({\rm wave \,1})$. In a minimal scenario with two
scalar triplets $T_{1,2}$, with $\mathbf{Y}^{T_1}\simeq \mathbf{Y}^{T_2}\sim
y$ and $M_1\ll M_2$, the condition for $\epsilon_a^{\alpha\beta}({\rm wave
\,2}) \gg \epsilon_a^{\alpha\beta}({\rm wave \,1})$ would roughly be
\begin{equation}\label{cond}
y^2\gg \frac{M_2}{M_1}\lambda_1 \lambda_2\,.
\end{equation}
This can be easily achieved if the triplets couple strongly to leptons but
very weakly to the SM Higgs doublet. Ultimately, if one of the triplets
couples mainly to the leptons (thus, not giving any contribution to neutrino
masses) then $\epsilon_a^{\alpha\beta}({\rm wave \,1})\simeq0$ and
$\epsilon_a^{\alpha\beta}({\rm wave \,2})$ is the only contribution for the
CP asymmetries. To illustrate this, let us consider a simple example with two
triplets $T_{1,2}$ of masses $M_{1,2}$. We assume
\begin{equation}\label{Ydeltas}
\mathbf{Y}^{T_1}\simeq \frac{M_1\mathbf{m}_\nu}{2\lambda_1v^2}\;,\; \mathbf{Y}^{T_2}=
\mathbf{K}\mathbf{Y}^{T_1}\mathbf{K}\,,
\end{equation}
with $\mathbf{K}={\rm diag}(e^{i\pi/2},1,1)$. The above approximation for
$\mathbf{Y}^{T_1}$ is valid provided that $\lambda_1 M_2 \gg \lambda_2 M_1$.
The effective neutrino mass matrix is constructed from
$\mathbf{m}_\nu=\mathbf{U}^\ast {\rm diag}(m_1,m_2,m_3) \mathbf{U}^\dag$,
where $\mathbf{U}$ is the PMNS lepton mixing matrix in the standard PDG
parametrization and $m_i$ are the neutrino masses. We take the best-fit
values from the latest global analysis of all neutrino oscillation
data\cite{Tortola:2012te,GonzalezGarcia:2012sz} and consider a hierarchical
neutrino mass spectrum with $m_1\simeq0$. We also assume maximal Dirac-type
CP violation, i.e. the phase $\delta=\pi/2$, and neglect any Majorana-type CP
violation. As for the high-energy parameters, we choose
$\lambda_1=10\lambda_2=5\times 10^{-6}$ and take $M_2=10 M_1=10^{10}\,{\rm
GeV}$ (to ensure that leptogenesis takes place within the flavoured regime).
In this case, $\epsilon_1^{\alpha\beta}({\rm wave \,1})\simeq 0$ (for all
$\alpha,\,\beta=e, \mu, \tau$), while
\begin{equation}\label{CP3matrix}
\epsilon_1^{\alpha\beta}({\rm wave \,2})\simeq\left(\begin{array}{ccc}
-0.02 &\,-5.62 &\, -6.92 \\
-5.62 &\,\,\,\, 5.84 &\,\,\,\, 4.59\\
-6.92 &\,\,\,\, 4.59 &\,\,\,\, 10.08
\end{array}\right)\times 10^{-7}\,,
\end{equation}
which is sufficiently large to give a sizeable contribution to the baryon
asymmetry.

\section{Flavoured Boltzmann equations}

The final asymmetry crucially depends on the efficiency of leptogenesis,
which is dictated by the solution of the relevant Boltzmann equations in the
flavoured regime. Before discussing these equations in detail, it is worth
commenting on some general features that are present in the unflavoured
regime. It has been shown\cite{Hambye:2005tk} that, for unflavoured
leptogenesis, the efficiency is maximal when either $\mathcal{B}^\ell_a \ll
\mathcal{B}^\phi_a$ or $\mathcal{B}^\ell_a \gg \mathcal{B}^\phi_a$ (for a
recent analysis see Ref.~\cite{Hambye:2012fh}). This can be easily understood
if one recalls that, in the type II seesaw, lepton number is violated only if
both decay channels of the scalar triplet (i.e. to two leptons and to two
Higgs scalars) are present. Thus, even when the total decay rate
$\Gamma_{T_a}$ and the gauge scattering rates are much larger than the Hubble
rate, if either the decay rate to leptons or to Higgs doublets is out of
thermal equilibrium (i.e. $\mathcal{B}^\ell_a \ll 1$ or $\mathcal{B}^\phi_a
\ll 1$), lepton number is not erased by the corresponding inverse decays, and
there is no suppression of the leptogenesis efficiency. As we shall see
below, some of these features remain valid in the flavoured regime.

We restrict our analysis to a minimal scenario of two scalar triplets
$T_{1,2}$, with $T_1$ lighter that $T_2$, so that the effects of
lepton-number violating processes due to $T_2$ can be safely neglected. In
general, the lepton asymmetry produced in the $T_2$ decays will be washed out
by the interactions of $T_1$.  We denote by $n_x$ the number density of the
particle $x$, and define its comoving number density $Y_x=n_x/s$, where
$s=(2\pi^2/45)\, g_\ast T^3$ is the total entropy density at leptogenesis
temperatures ($g_\ast=106.75$). We also define the comoving asymmetries
$\Delta_x=Y_x-Y_{\bar{x}}$ and denote by $\Sigma_T=Y_{T_1}+Y_{\bar{T}_1}$ the
total triplet density.

The relevant Boltzmann equations, which describe the evolution of the
asymmetries $\Delta_T$, $\Delta_\phi$ and $\Delta_{\ell_\alpha}\,(\alpha=e,
\mu, \tau)$ as functions of $z=M_1/T$, read as~\footnote{Since scalar
triplets are not self-conjugated states, a triplet-antitriplet asymmetry
$\Delta_{T}$ is generated and a Boltzmann equation for this asymmetry must be
included.}

\begin{subequations} \label{Boltzeqs}
\begin{align}
s z H(z) \,\frac{d\Sigma_T}{dz}=&-\left(\frac{\Sigma_T}{\Sigma_T^{eq}}-1\right)\gamma_D
-2\left(\frac{\Sigma^2_T}{\Sigma_T^{eq^2}}-1\right)\gamma_A\,,\label{BE1}\\
s z H(z) \,\frac{d\Delta_T}{dz}=&-\gamma_D\left(\frac{\Delta_T}{\Sigma_T^{eq}}
+\sum_{\alpha,\beta}\mathcal{B}_1^{\alpha\beta}\frac{\Delta_{\ell_\alpha}}{Y_\ell^{eq}}
-\mathcal{B}^\phi_1 \frac{\Delta_\phi}{Y_\phi^{eq}}\right)\,,\label{BE2}\\
s z H(z) \,\frac{d\Delta_\phi}{dz}=&\sum_{\alpha,\beta}X_{\alpha\beta}
-2\mathcal{B}^\phi_1\gamma_D\left(\frac{\Delta_\phi}{Y_\phi^{eq}}
-\frac{\Delta_T}{\Sigma_T^{eq}}\right)\,,\label{BE3}\\
s z H(z) \,\frac{d\Delta_{\ell_\alpha}}{dz}=&\sum_{\beta}\left[X_{\alpha\beta}
-2\mathcal{B}_1^{\alpha\beta}\gamma_D\left(\frac{\Delta_T}{\Sigma_T^{eq}}
+\frac{\Delta_{\ell_\alpha}+\Delta_{\ell_\beta}}{2Y_\ell^{eq}}\right)\right]\,,\label{BE4}
\end{align}
\end{subequations}
where
\begin{align}
\begin{split}
X_{\alpha\beta}=&\left(\frac{\Sigma_T}{\Sigma_T^{eq}}-1\right)
\gamma_D\epsilon_1^{\alpha\beta}
+2\gamma_D\left(\mathcal{B}^\ell_1\,\epsilon_1^{\alpha\beta}
-\mathcal{B}_1^{\alpha\beta}\,\epsilon_1\right)\\
&-\left(2\frac{\Delta_\phi}{Y_\phi^{eq}}+\frac{\Delta_{\ell_\alpha}
+\Delta_{\ell_\beta}}{Y_\ell^{eq}}\right)\left(2\gamma_{\ell_\alpha
\ell_\beta}^{\prime\bar{\phi}\bar{\phi}}
+\gamma_{\ell_\alpha\phi}^{\bar{\ell}_\beta\bar{\phi}}\right).
\end{split}
\end{align}
In the above equations,
\begin{equation}
H(z)=\frac{H_0(M_1)}{z^2}\,,\quad
H_0(T)=\sqrt{\frac{4\pi^3}{45}g_\ast}\,\frac{T^2}{m_{P}}
\end{equation}
is the Hubble constant at temperature $T$, $m_{P}=1.22 \times 10^{19}$~GeV is
the Planck mass, and the suffix $eq$ denotes equilibrium values. We have
\begin{equation}
Y_T^{eq}=\frac{45\, g_T}{4\pi^4 g_\ast}z^2 K_2(z)\,,\quad
Y_{\ell}^{eq}=\frac{3}{4}\frac{45\,\zeta(3)g_\ell}{2\pi^4 g_\ast}\,,\quad
Y_\phi^{eq}=\frac{45\,\zeta(3)g_\phi}{2\pi^4 g_\ast}\,,
\end{equation}
where $g_x$ are the degrees of freedom of the particle ($g_T=1$ for each
triplet component, $g_\ell=2$ and $g_\phi=2$), $\zeta(3)\simeq 1.202$ and
$K_i(z)$ are the modified Bessel functions. Finally, the relevant interaction
densities contributing to leptogenesis are the decays and inverse decays,
given by the standard expression
\begin{equation}
\gamma_D=s\,\Gamma_{T_1}\,\Sigma_T^{eq}\,\frac{K_1(z)}{K_2(z)}\,,
\end{equation}
gauge scattering processes of the triplets, approximately given by the s-wave
contribution
\begin{equation}
\gamma_A=\frac{M_1 T^3 e^{-2M_1/T}}{64\pi^4}(9 g^4 + 12g^2g_Y^2 + 3g_Y^4)
\left(1+\frac{3T}{4M_1}\right),
\end{equation}
and $\Delta L=2$ scattering processes due to $\ell_\alpha \ell_\beta
\leftrightarrow \bar{\phi} \bar{\phi}$ and $\ell_\alpha \phi \leftrightarrow
\bar{\ell}_\beta \bar{\phi}$ generated in $s$-channel and $t$-channel,
respectively. For these processes, the reaction densities are obtained as
\begin{equation}
\gamma=\frac{T}{64\pi^4}\int_{0}^\infty ds\, s^{1/2}
K_1(\sqrt{s}/T)\,\hat{\sigma}(s),
\end{equation}
where the reduced cross sections are
\begin{subequations}
\begin{align}
\hat{\sigma}_{\ell_\alpha \ell_\beta}^{\bar{\phi}\bar{\phi}}
=\frac{3 x |\lambda_1|^2 |\mathbf{Y}^{T_1}_{\alpha\beta}|^2}{2 \pi}
\left[\frac{1}{(1-x)^2+(\Gamma_{T_1}/M_1)^2}\right]\,,\\
\hat{\sigma}_{\ell_\alpha\phi}^{\bar{\ell}_\beta\bar{\phi}}
=\frac{6 |\lambda_1|^2 |\mathbf{Y}^{T_1}_{\alpha\beta}|^2}{\pi}
\left[-\frac{1}{1+x}+\frac{\ln(1+x)}{x}\right]\,,
\end{align}
\end{subequations}
and $x=s/M_1^2$. We recall that in the Boltzmann equations the term due to
on-shell triplet exchange must be subtracted from $\gamma_{\ell_\alpha
\ell_\beta}^{\bar{\phi}\bar{\phi}}$. This procedure leads to the subtracted
reaction density
\begin{equation}
\gamma^\prime_{\alpha\beta}=\gamma_{\alpha\beta}
-\frac{1}{2}\, \mathcal{B}_1^{\alpha\beta}\mathcal{B}^\phi_1\gamma_D,
\end{equation}
where $\mathcal{B}_1^{\alpha\beta}= \Gamma(T^\dagger_1\rightarrow
\ell_{\alpha} \ell_{\beta})/\Gamma_{T_1}$.

The Boltzmann equations given in Eqs.~\eqref{Boltzeqs} have been obtained
following the standard general procedure, as described e.g. in
Ref.~\cite{Davidson:2008bu}. These equations contain the relevant processes
contributing to triplet leptogenesis and share the same structure of those in
the unflavoured regime\cite{Hambye:2005tk}. The main difference resides in
the fact that quantities that depend on the lepton flavours are now treated
independently.

\begin{figure}[t]
\begin{center}
\includegraphics[width=10.5cm]{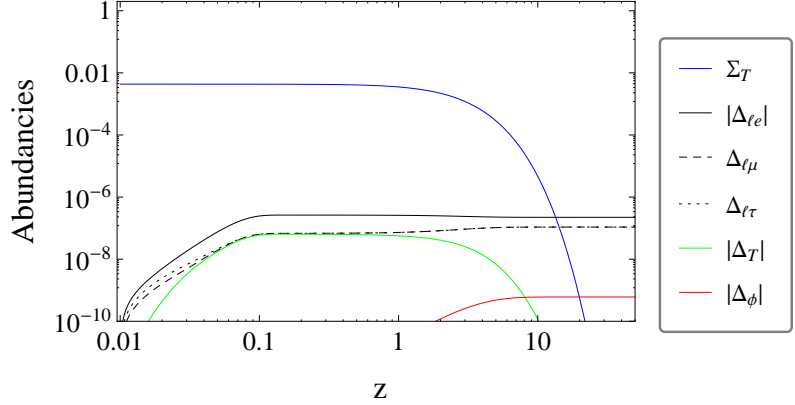}
\end{center}
\caption{\label{fig2} Evolution of asymmetries for the example case in which
the flavoured CP asymmetries are given by Eq.~\eqref{CP3matrix}, $M_2 = 10
M_1 = 10^{10}$~GeV and $\lambda_1=10\lambda_2=5\times 10^{-6}$. The final
baryon asymmetry is $\Delta_B  \simeq 8 \times 10^{-8}$.}
\end{figure}

We integrate the system of equations~\eqref{Boltzeqs} with the following
initial conditions: $\Sigma_T(z \ll 1)=\Sigma_T^{eq}(z\ll 1)$,\, $\Delta_T(z
\ll 1)=0$,\, $\Delta_\phi(z \ll 1)=0$, and $\Delta_{\ell_\alpha}(z \ll 1)=0$.
To demonstrate how a large leptogenesis efficiency may arise due to the novel
CP asymmetry contribution given in Eq.~\eqref{CPasymmIIflavour2}, we consider
the example case presented at the end of the previous section. In
Fig.~\ref{fig2} we plot the evolution of the asymmetries $\Delta_T$,\,
$\Delta_\phi$, and $\Delta_{\ell_\alpha}$ in the three-flavoured regime, with
the CP asymmetries given by the matrix~\eqref{CP3matrix}, $M_2 = 10 M_1 =
10^{10}$~GeV and $\lambda_1=10\lambda_2=5\times 10^{-6}$. As can be seen from
the figure, large leptonic asymmetries develop,
\begin{align}
\Delta_{\ell e}\simeq -2.23\times10^{-7},
\quad \Delta_{\ell \mu}\simeq \Delta_{\ell \tau} \simeq 1.08\times10^{-7},
\end{align}
and a sizable baryon asymmetry can be generated. Indeed, at temperatures
below $10^9$~GeV, the final baryon asymmetry $\Delta_B$ can be estimated as
\begin{equation}
\Delta_B=3\, c_{\rm sph} \sum_{\alpha,\beta}A^{-1}_{\alpha\beta}\,\Delta_{\ell_\beta},
\end{equation}
where $c_{\rm sph}=28/79$ is the sphaleron conversion factor, and the matrix
$A$ is given by\cite{Nardi:2006fx}
\begin{equation}
A=
\begin{pmatrix}
-151/179&20/179&20/179\\
25/358&-344/537&14/537\\
25/358&14/537&-344/537
\end{pmatrix}.
\end{equation}
For the example under consideration, we obtain $\Delta_B  \simeq 8 \times
10^{-8}$.

It is worthwhile to comment on the efficiency of leptogenesis in this case.
Defining the flavoured efficiency factors,
\begin{align}
\eta_\alpha = \frac{|\Delta_{\ell \alpha}|}{|\sum_\beta \epsilon_1^{\alpha\beta}|},
\end{align}
we estimate $\eta_e \simeq 0.18, \quad \eta_\mu \simeq 0.22, \quad \eta_\tau
\simeq 0.14$, i.e. the efficiency in all flavours is large. This stems from
the fact that the decay of the scalar triplet into two Higgs doublets is
strongly out of equilibrium, since $\mathcal{B}_1^\phi\, \Gamma_{T_1} \ll H$.
Notice also that the values of the efficiency parameters for the various
lepton flavours differ from each other. This is required to guarantee that
the total lepton asymmetry does not vanish. The crucial point here is that
the strength of the washout parameters is mainly controlled by the Yukawa
couplings $\mathbf{Y}^{T_1}_{\alpha\beta}$, which are in general different
for each lepton flavour. These couplings enter directly into the Boltzmann
equations through the leptonic branching ratios $\mathcal{B}_1^{\alpha\beta}$
(cf. Eqs.~\eqref{BE2} and \eqref{BE4}). Were all these branching ratios
equal, then the efficiency of leptogenesis would be the same in all flavours,
leading to a vanishing total lepton asymmetry\,\footnote{We remark that, by
summing over the lepton flavours $\alpha$ in Eq.~\eqref{BE4}, one does not
recover the unflavoured Boltzmann equation for the total lepton
asymmetry\cite{Hambye:2005tk}, unless all $\mathcal{B}_1^{\alpha\beta}$ are
equal.}.

\begin{figure}[t]
\begin{center}
\includegraphics[width=9cm]{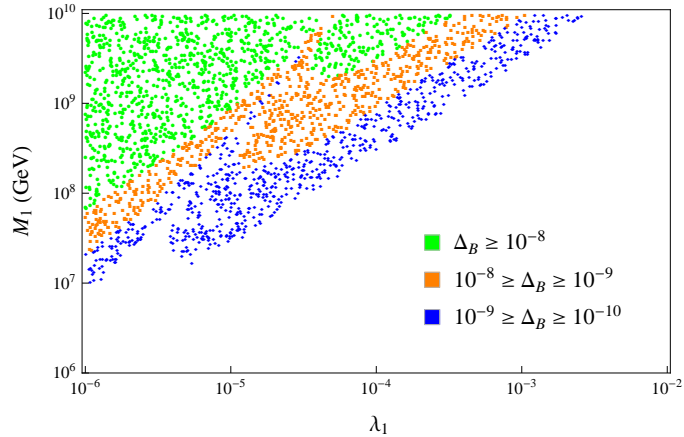}
\end{center}
\caption{\label{fig3} Baryon asymmetry in the ($\lambda_1,M_1$) parameter
space, for the example case with the flavoured CP asymmetries of
Eq.~\eqref{CP3matrix}, $M_2 = 10 M_1$ and $\lambda_1=10\lambda_2$.}
\end{figure}

We have also randomly varied the triplet mass $M_1$ and the coupling
$\lambda_1$, but keeping the relations $M_2 = 10 M_1$ and
$\lambda_1=10\lambda_2$. The results are presented in Fig.~\ref{fig3}, which
clearly shows that there is a large region of the parameter space where
flavoured type II seesaw leptogenesis is efficient and leads to a viable
baryon asymmetry, exclusively dominated by the CP asymmetry coming from the
one-loop diagram (1c) in Fig.~\ref{fig1}.

\section{Conclusions}

The class of scenarios where the total unflavoured lepton asymmetry is zero
($\epsilon_1=0$), while its individual flavour contributions are not, is
known as purely flavoured leptogenesis (PFL)\cite{AristizabalSierra:2009bh}.
These scenarios have been studied in type-I seesaw leptogenesis, and it has
been shown that, in order to get a sizable contribution from the Boltzmann
equations, the various flavour projections have to be different. In the case
where these projections are of the same order for each flavour, known as
lepton flavour equilibration, PFL may become
ineffective\cite{AristizabalSierra:2009mq}. This is however a model-dependent
conclusion which cannot be generalized to all PFL
models\cite{Pascoli:2006ie,Branco:2006ce,AristizabalSierra:2009bh,GonzalezGarcia:2009qd}.
In the context of type-II seesaw PFL, the relevance of lepton flavour
equilibration and flavour-dependent washout processes through the study of
the Boltzmann equations has not been addressed yet. The example case
presented in this work is a first step in this direction.

From a theoretical viewpoint, a simple way to suppress the contribution of
diagram (b) to the leptonic CP asymmetry with respect to the one of diagram
(c) (cf. Fig.~\ref{fig1}) is by imposing some symmetry that forbids the
trilinear terms $\mu_a\tilde{\phi}^TT_a\tilde{\phi}$. Clearly, in this case,
no effective neutrino mass term can be generated. To implement the type II
seesaw mechanism the symmetry should be softly broken. A well-known example
is the soft breaking of lepton number $L$. For instance, considering a
Lagrangian invariant under $U(1)_L$ with the symmetry transformations
$\ell_{L}\rightarrow e^{i\alpha_L}\ell_L\,,\,e_{R}\rightarrow
e^{i\alpha_L}e_R\,,\, T_a\rightarrow e^{-2i\alpha_L}T_a\,,\, \phi\rightarrow
\phi\,$, only the lepton-number conserving loop diagram~(c) is allowed.
Neutrino masses can then be generated by an explicit soft breaking of the
lepton number, or by the spontaneous breaking induced by the vacuum
expectation values of additional scalar fields, $\mu_a \sim \langle \eta
\rangle$. A drawback in the latter scenario is the appearance of pseudo
Nambu-Goldstone bosons (Majorons) after the spontaneous breaking of the
global lepton number, which are tightly constrained by experimental searches.
This problem can be avoided by replacing the continuous symmetry by a
discrete one; for example, by requiring the Lagrangian to be invariant under
the $Z_3$ transformations $\{\ell_{L},e_{R},T_a\} \rightarrow e^{i2\pi/3}
\{\ell_{L},e_{R},T_a\},\, \phi \rightarrow \phi$.

To conclude, in this work we have studied the relevance of flavoured effects
on the leptonic CP asymmetries generated in purely type II seesaw scenarios,
where leptogenesis is implemented through the out-of-equilibrium decays of
scalar triplets. We have found a novel contribution to the flavoured
leptogenesis asymmetries coming from the wave-function renormalization
correction, with leptons running inside the loops. This contribution
conserves total lepton number but violates lepton flavour and, therefore,
does not vanish in the flavoured leptogenesis regime at temperatures $T
\lesssim 10^{12}$~GeV. We have also briefly discussed some possible scenarios
in which such contribution may dominate the cosmological baryon asymmetry.

\section*{Acknowledgments}
This work was supported by Portuguese national funds through FCT - Funda\c{c}\~{a}o
para a Ciencia e Tecnologia, through the project PEst-OE/FIS/UI0777/2011 and
the project CERN/FP/116328/2010. The work of H.S. was funded by the European
FEDER, Spanish MINECO, under the grant FPA2011-23596.

\end{document}